\documentclass[a4paper,twocolumn,english,aps,prl,floatfix,showpacs]{revtex4}
\usepackage[T1]{fontenc}
\usepackage{amsmath}
\usepackage{graphics}
\usepackage{graphicx}
\usepackage{color}

\begin{document}

\title{High fidelity transfer of an arbitrary quantum state between harmonic oscillators}

\author{K. Jahne$^{1,2}$, B. Yurke$^3$, and U. Gavish$^{1,2}$}

\affiliation{1. Institute for Theoretical Physics, University of
Innsbruck, Innsbruck A-6020, Austria\\
2. Institute for Quantum Optics and Quantum Information of the
Austrian Academy of Sciences, A-6020 Innsbruck, Austria\\
3. Bell Laboratories, Lucent Technologies, Murray Hill, NJ 07974}
\begin{abstract}
It is shown that by switching a specific time-dependent interaction between a harmonic
oscillator and a transmission line (a waveguide, an optical fiber, etc.) the quantum state
of the oscillator can be transferred into that of another oscillator coupled to the
distant other end of the line, with a fidelity that is \emph{independent}
of the initial state of both oscillators.
For a transfer time $T,$ the fidelity approaches 1 exponentially in $\gamma T$ where
$\gamma$ is a characteristic damping rate.  Hence, a good fidelity is
achieved even for a transfer time of  a few damping times.
Some implementations are discussed.
\end{abstract}

\date{\today}

\maketitle

A \emph{state-transfer} between two \emph{identical} distant systems (say, two nodes of a quantum network)
 is a process in which at time $t=T$ the second system obtains
the same quantum state that the first one had at time $t=0.$ The systems between which
the state is transferred may include, for example, atoms in a cavity
\cite{Cirac},\cite{Pellizzari}, spin or flux qubits\cite{bose},\cite{bruder}. The need for a
fast and reliable state-transfer in quantum computers and quantum networks is posing
questions concerning the practical and principle limitations on the speed and reliability
of such a communication form  (in addition to the obvious limitations related to the
speed of light). These questions were considered mainly \cite{bose}-\cite{de Pasquale},
though not only  \cite{Yung networks}-\cite{MIT}, in condensed matter systems where the transferred
states belong to a finite dimensional Hilbert space (e.g. that of 2 or 3 level systems
and unlike the state of a harmonic oscillator).

Here we consider a state transfer between two identical harmonic
 oscillators coupled to opposite ends of a transmission line. We show that by performing
a single operation of properly chosen time-dependent switching of the coupling (denoted
by $\gamma(t)$) between one of the oscillators and the transmission line, a state
transfer is achieved with an arbitrary high fidelity which is \emph{independent} of the initial states
of the two oscillators.

 We start by analyzing the case where a single oscillator (labeled $i$) is coupled to a transmission
line  with a constant coupling, $\gamma_i.$
 Assuming that $\gamma_i$ (which also characterizes the damping of the oscillator)
is much smaller than the oscillator frequency $\omega_0$, one is allowed to use the
slowly varying envelope, Markov, and rotating-wave approximations. In a frame rotating at the oscillator frequency,
 the Heisenberg equations of motion are then given by (see, e.g., Eqs. (7.15) and (7.18) in
Ref. \cite{Walls}):
\begin{eqnarray}
\label{1a}
\frac  {d a_i }  { d t}+\gamma_i a_i=\sqrt{2\gamma_i}b_{i,in}(t),~~~~~i=1,2
\end{eqnarray}
and
\begin{eqnarray}
\label{2a}
\sqrt{2\gamma_i}a_i=b_{i,in}(t)+b_{i,out}(t).
\end{eqnarray}
$a_i(t)$ annihilates a mode of the oscillator $i$ and satisfies
$[a_i(t),a_i^{\dagger } (t)]=1.$ $b_{i,in/out}(t)$ is an operator related
to the incoming/outgoing fields in the transmission line by \cite{Denker}
\begin{eqnarray}
\label{3a}
b_{i,\alpha}(t)=\frac{1}{\sqrt{2\pi}}\int\limits_{-\infty}^{\infty}d\omega b_{i,\alpha}(\omega )e^{-i\omega t}
~~~~~\alpha=in,out\end{eqnarray}
where $b_{i,\alpha}(\omega )$ annihilates a transmission line mode
propagating towards (for $\alpha=in$) or away from (for $\alpha=out$) the oscillator $i$, and
\begin{eqnarray}
\label{4a}
~[b_{i,\alpha}(t),b_{i,\alpha}^{\dagger } (t')]=\delta(t-t'),~~~[b_{i,\alpha}(t),b_{i,\alpha}(t')]=0.
\end{eqnarray}
$\gamma_i$ in Eq.(\ref{1a}) is  the line-oscillator coupling strength and thus appears both
in the damping term on the right of Eq. (\ref{1a}) and the driving term on
the left. Eq. (\ref{2a}) is a boundary condition matching the value of the
oscillator field with that of the transmission line at their interface.
 Eqs. (\ref{1a}) and (\ref{2a}) provide a rather general description of
a weakly-damped harmonic oscillator (a cavity resonator or an LC circuit)
  coupled to a continuum of propagating harmonic modes such as those
of free space or a waveguide such as an optical fiber or an electrical
transmission line.
Specific examples of circuit implementations leading to Eqs. (\ref{1a}) and
(\ref{2a}) are discussed below.
Their solution in terms of the incoming modes is
\begin{eqnarray}
\label{5c} a_i(t)=e^{-\gamma_i t}a_i(0)+\sqrt{2\gamma_i}\int_0^{t}dt'
e^{-\gamma_i(t-t')}b_{i,in}(t').
\end{eqnarray}
Now we proceed to analyze two oscillators coupled to both ends of a transmission line. We
consider the oscillators to be 'cascaded' i.e. the modes propagating away from
oscillator 1 are propagating into oscillator 2 while the modes propagating away from
oscillator 2 are directed into a termination which functions as a black-body absorber
  (preventing backward scattering of the modes)
at a temperature of absolute zero. This can be achieved
by employing an isolator, such as the terminated three port circulator shown in
Fig. 1, in the transmission line.  Denoting
 the propagation time of the signal along the line by $\tau,$ one therefore has
$b_{1,out}(t)=b_{2,in}(t+\tau)$. In order to express our results in a form independent of
the transmission line length we shift the time origin of oscillator 2  by $\tau$ and thus
write
\begin{eqnarray}
\label{6c}
b_{1,out}(t)=b_{2,in}(t).
\end{eqnarray}
The equations of motion are thus given by Eq. (\ref{6c}) and Eqs. (\ref{1a}), (\ref{2a}) for
$i=1,2.$ Their solution is:
\begin{eqnarray}
\label{7c}
a_2(t)=e^{-\gamma t} a_2(0)+2\gamma t e^{-\gamma t}a_1(0)\nonumber  \\
-\sqrt{2\gamma}e^{-\gamma t}\int\limits_0^{t}dt' e^{\gamma t'}b_{1,in}(t')\nonumber  \\
+(2\gamma)^{3/2}e^{-\gamma t}\int\limits_0^{t}dt'\int\limits_0^{t'}dt'' e^{\gamma
t''}b_{1,in}(t'').
\end{eqnarray}
 The fidelity, $F(t),$ of the state transfer is defined as the coefficient in front of $a_1(0).$
 Here
 \begin{eqnarray}
\label{8c} F(t)=2\gamma t e^{-\gamma t}.
 \end{eqnarray}
At $t=\frac{1}{\gamma}$ it obtains its maximal value $F_{max}\approx 0.736.$
 Note that the above definition implies that the fidelity is \emph{independent}
of the initial states of the oscillators. This advantage over the standard \emph{state-dependent}
definition of fidelity \cite{Nielsen} (as an overlap between the final state and the desired transferred one)
 is made possible by the linearity of the \emph{operator} relation,
Eq.(\ref{7c}).

If a time-dependent switch is inserted between oscillator 1 and the line, the system is
still governed by Eqs. (\ref{1a}), (\ref{2a}) and (\ref{6c}), except that
$\gamma_1(t)$ is now time-dependent while $\gamma_2=\gamma$ remains constant. The
solution is given by
\begin{eqnarray}
\label{9c}
a_1(t)=a_1(0)e^{-\int\limits_0^t dt' \gamma_1(t')}~~~~~~~~~~~~~~~~~~~~\nonumber  \\
+\int\limits_0^t dt' \sqrt{2\gamma_1(t')}e^{-\int\limits_{t'}^t
dt''\gamma_1(t'')}b_{1,in}(t')
\end{eqnarray}
and
\begin{eqnarray}
\label{10c}
a_2(t)=e^{-\gamma t} a_2(0)~~~~~~~~~~~~~~~~~~~~~~~~~~~~~~~~~~~~~~~~\nonumber  \\
+2\sqrt{\gamma}\int\limits_0^t dt' e^{-\gamma (t-t')}
\sqrt{\gamma_1(t')}e^{-\int\limits_0^{t'}dt''\gamma_1(t'')}a_1(0)\nonumber  \\
+2\sqrt{2\gamma}
\int\limits_0^t dt' e^{-\gamma (t-t')}\sqrt{\gamma_1(t')}\times \nonumber  \\
\int_0^{t'}dt''\sqrt{\gamma_1(t'')}
e^{-\int\limits_{t''}^{t'}dt'''\gamma_1(t''')}b_{1,in}(t'')\nonumber  \\
-\sqrt{2\gamma}\int\limits_0^t dt' b_{1,in}(t')e^{-\gamma (t-t')}.
\end{eqnarray}
%
The (once again state-independent) fidelity is therefore
\begin{eqnarray}
\label{11c} F(t)=2\sqrt{\gamma}\int\limits_0^t dt' e^{-\gamma (t-t')}
\sqrt{\gamma_1(t')}e^{-\int\limits_0^{t'}dt''\gamma_1(t'')}.
\end{eqnarray}
It remains to find $\gamma_1(t)$ that maximizes $F(t).$ To this end we denote
$\int_0^t dt' \gamma_1(t') =G(t)$ and rewrite Eq. (\ref{11c}) as
\begin{eqnarray}
\label{12c}  F\left[G,\frac  {d G }  { d t},T\right]=
2\sqrt{\gamma}\int\limits_0^T dt' e^{-\gamma (T-t')} \sqrt{\frac  {d G }  { d
t'}}e^{-G(t')}.
\end{eqnarray}
The Euler-Lagrange equation corresponding to the extremum of the above integral (expressed
in terms of $\gamma_1=\frac  {d G }  { d  t}$)
is
\begin{eqnarray}
\label{13c}
2(\gamma_1)^2+2\gamma\gamma_1-\frac {d \gamma_1 }  {  d  t}=0.
\end{eqnarray}
This nonlinear equation is exactly solvable. The solution is
$\gamma_1(t)=\gamma\frac{1}{c_0e^{-2\gamma t}-1},$ $c_0$ is constant.
Denoting $c_0=e^{2\gamma T}$ (with $T\ge t$, so that
$\gamma_1>0$) one obtains
\begin{eqnarray}
\label{15c}
\gamma_1(t)=\gamma\frac{1}{e^{2\gamma (T-t)}-1}
\end{eqnarray}
and then from Eq. (\ref{11c})
\begin{eqnarray}
\label{16c} F(t)=2\frac{\sinh(\gamma t)}{\sqrt{e^{2\gamma T}-1}}.
\end{eqnarray}
Since $\sinh(t)$ is monotonic and $T \ge t$ the highest fidelity is obtained when $t=T$
and therefore we can identify $T$ as the total switching time and write
\begin{eqnarray}
\label{17c}
F(T)=\sqrt{1-e^{-2\gamma T}}.
\end{eqnarray}
Eqs. (\ref{15c}) and (\ref{17c})  (with its generalization Eq.(\ref{22d}))  are our main results. They demonstrate that
a state-transfer with arbitrary good fidelity can be achieved between two identical
oscillators, and that, with the proper choice of the switching function Eq.
(\ref{15c}), the fidelity approaches unity exponentially fast and therefore does not
require the operation time to be much longer than the damping time. For example,
Eq. (\ref{17c}) implies that $\gamma T>5$ suffices
 to achieve a fidelity greater than $1-10^{-4}.$

The above results relied on the small damping assumption,
 while $\gamma_1$ in Eq.(\ref{15c}) is singular near the point
$t=T$ where $\gamma_1(t)\approx  \frac{1}{2(T-t)}.$ In particular, the small damping
condition $\gamma_1\ll \omega _0$ breaks down near this point. Thus, we should
specify under what condition cutting off $\gamma_1(t)$ at time $T-\Delta t$ (where
$\gamma_1(t)\ll \omega _0$ still holds) will have negligible effect on the
fidelity. Expanding Eq.(\ref{16c}) for $t=T-\Delta t$ with $\gamma \Delta t  \ll 1$ (and $e^{-2\gamma T}\ll 1$)
yields
\begin{eqnarray}
\label{18c} 1-F=\frac{1}{2}e^{-2\gamma T}+\gamma \Delta t +O(\gamma \Delta
t)^2.
\end{eqnarray}
We see that an additional requirement of the form $\gamma \Delta  t \ll  1-F$ appears
 to ensure high fidelity. Recalling the asymptotic behavior of $\gamma_1(t)$ near
$t=T$ and the small damping condition $\gamma_i \ll \omega _0,$
 one obtains a sufficient condition for achieving arbitrary high fidelity (for long enough $T$)
\begin{eqnarray}
\label{19c} \omega _0\gg \gamma_1^{max}\gg \frac{\gamma_2}{1-F},
\end{eqnarray}
where $\gamma_1^{max}=\frac{1}{2\Delta t}.$
In terms of oscillator quality factors ($Q = \omega_0/\gamma$)
this implies
\begin{eqnarray}
\label{20c}
 (1-F) Q_2\gg Q_1^{min}\gg 1.
\end{eqnarray}

 To include losses in the oscillators and the line we introduce an oscillator damping-rate $\gamma'$
and a finite transmission coefficient through the line
$\eta\le 1$ by replacing Eqs.(\ref{1a}), (\ref{2a}) and (\ref{6c}) with
$\frac  {d a_i }{ d t}+(\gamma_i+\gamma') a_i=\sqrt{2\gamma_i}b_{i,in}(t)+\sqrt{2\gamma'}b'_{i,in}(t),$
$a_i=\frac{1}{\sqrt{2\gamma_i}}(b_{i,in}(t)+b_{i,out}(t))+\frac{1}{\sqrt{2\gamma'}}(b'_{i,in}(t)+b'_{i,out}(t)),$
(where $b'_{i,in/out}(t)$ are the loss channel modes) and $b_{2,in}(t)=\eta^{1/2}b_{1,out}(t)$.
One then obtains:
 \begin{eqnarray}
\label{21d} F(t)=2\eta^{1/2} e^{-\gamma' t}\frac{\sinh(\gamma t)}{\sqrt{e^{2\gamma T}-1}},
\end{eqnarray}
\begin{eqnarray}
\label{22d}
F(T)=\eta^{1/2}e^{-\gamma'T}\sqrt{1-e^{-2\gamma T}},~~~~\gamma'<\gamma,
\end{eqnarray}
and a criterion for the losses being small (as far as the fidelity is concerned):
  $1-\eta\ll1$ and $\gamma' T\ll 1.$

We now turn to specific optical and electronic realizations of the system
analyzed above. Before discussing the implementation of the time-dependent
switching we show how a single LC circuit coupled
to a line with constant coupling evolves according to Eqs.(\ref{1a}) and (\ref{2a}).
The generalization to the case where each of the oscillators is a collective mode of a
combination of several LC circuits presents no special difficulties.

Considering an ideal $LC$ circuit coupled to a transmission line
 with a (real) impedance $R,$ the Heisenberg equations of motions (see e.g., Eq. (3.14) in \cite{Denker}) are
\begin{eqnarray}
\label{1} \frac  {d ^2  x_i }{  d t^2}+2 \gamma_i\frac  {d   x_i}  {  d  t} +\omega
_0^2x_i=  4\gamma_i\frac  {d  x_{i,in} }  {  d  t},
\end{eqnarray}
and
\begin{eqnarray}
\label{2}
x_i=x_{i,in}+x_{i,out}.
\end{eqnarray}
If the line, capacitor and inductor are in series, then $x_i$ in Eq.
(\ref{1}) stands for the charge on the capacitor and $ \gamma_i=\frac{R}{2L}$. If they
are in parallel, then $x_i$ stands for the voltage drop over the capacitor and
$\gamma_i=\frac{1}{2RC}.$ In both cases, $\gamma_i$ is the damping coefficient and since
the energy of the LC oscillator is transferred to the line, $\gamma_i$ also determines
the oscillator-line coupling strength. $x_{i,in}$ and $x_{i,out}$ are the transmission
line charge (for series circuit)  or voltage (for parallel circuit)
 excitation modes propagating towards and away from the oscillator at the point where it joins the
transmission line \cite{Denker}:
\begin{eqnarray}
\label{3}
x_{i,\alpha}(t)=N_R\int\limits_0^{\infty}d\omega \omega ^{-s/2}
a_{i,\alpha}(\omega )e^{-i\omega (t+q_{\alpha} \frac{y_i}{v})} ~ + h.c.
\end{eqnarray}
$v$ is the signal propagation speed along the line (assuming no dispersion).
$y_i$ is the position of the LC circuit. This position is well defined if the circuit is assumed to be 'lumped', i.e.
much smaller in size than the wave length of the transmission line excitation.
The $a_{i,out/in}$'s are the annihilation operators of the line excitations satisfying bosonic commutation relations.
$N_R=\sqrt{\frac{\hbar }{4\pi R}},~s=1$ for the charge  excitations
 and $N_R=\sqrt{\frac{\hbar R}{4\pi }},~s=-1$ for the voltage excitations.
$q_{in}=1,~q_{out}=-1.$

 Eq. (\ref{1}) describes a damped oscillator driven by the incoming
modes in Eq. (\ref{3}). With no damping and driving its
solution would have the usual form $x_i(t)=\Delta x_0 (a_ie^{-i\omega_0 t}+a_i^{\dagger }
e^{i\omega_0 t})$, where $\Delta x_0$ is the  ground state fluctuation
(for the charge in the series LC circuit $\Delta x_0=\sqrt{\frac{\hbar }{2\omega _0L}},$
for the voltage in the parallel LC circuit $\Delta x_0=\sqrt{\frac{\hbar\omega _0 }{2C}}$ ).
Assuming a small damping, $\gamma_i\ll \omega _0,$  we write
\begin{eqnarray}
\label{4} x_i(t)=\Delta x_0 \left(a_i(t)e^{-i\omega_0 t}+a_i^{\dagger }(t) e^{i\omega_0
t}\right),
\end{eqnarray}
where now $a_i(t)$  are slowly varying
 functions (still satisfying bosonic commutation relations).
 Neglecting derivatives which are not multiplied by a large parameter such as $\omega _0$
(slowly varying envelope approximation) we write
\begin{eqnarray}
\label{5} \frac  {d   x_i}  {  d  t}= -i\omega _0\Delta x_0  (a_i e^{-i
\omega_0 t}-a_i^{\dagger } e^{i \omega_0 t})
\end{eqnarray}
and
\begin{eqnarray}
\label{6} \frac {d ^2  x_i}  {d t^2}= -\omega _0^2 x_i-2i\omega _0\Delta x_0 \left(\frac
{d a_i }  { d  t}e^{-i\omega_0 t}-\frac  {d a_i^{\dagger } }  { d t} e^{i\omega_0
t}\right).
\end{eqnarray}
To separate the slow and the fast contribution from the transmission line modes we
approximate Eq. (\ref{3}) by
\begin{eqnarray}
\label{7} x_{i,\alpha}(t)= ~~~~~~~~~~~~~~~~~~~~~~~~~~~~~~~~~~~~~~~~~~~~~~~~~~~~~~~~~~\nonumber  \\
N_R e^{-i\omega
_0(t+q_{\alpha}\frac{y_i}{v})}\int\limits_{-\infty}^{\infty}d\omega \omega_0 ^{-s/2}
b_{i,\alpha}(\omega )e^{-i\omega (t+q_{\alpha}\frac{y_i}{v})}
 + h.c.
\end{eqnarray}
where $ b_{i,in}(\omega )\equiv a_{i,in}(\omega _0+\omega ).$  Eq. (\ref{7})
is a (Markov) approximation since the lower integration limit is extended to $-\infty .$
It is justified, because the assumption of small damping means that
 the oscillator responds only in a narrow frequency band around $\omega _0.$
Under these assumptions the  Fourier transforms of
$b_{i,\alpha}(\omega ),$ defined by  Eq. (\ref{3a}), satisfy
 Eq. (\ref{4a}) (for $\alpha=in,out$).
Setting $y_i=0$, making use of  Eqs. (\ref{4})-(\ref{7}) and (\ref{3a})
in Eqs. (\ref{1}) and (\ref{2}),
 and separating frequencies
 one obtains  Eqs. (\ref{1a}) and (\ref{2a}).

\begin{figure}
     \begin{center}
         \includegraphics[scale=0.31]{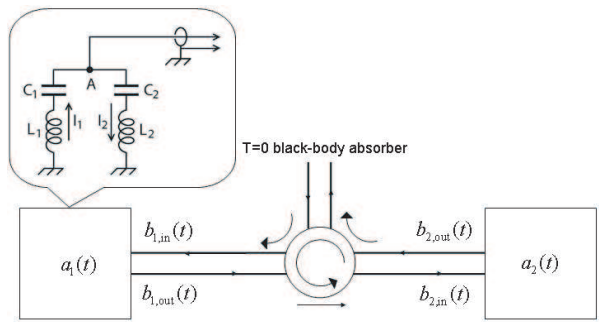}
     \end{center}
\caption{ Two identical harmonic oscillators $a_1$ and $a_2$ coupled to opposite ends of a transmission line.
An isolator (round object) placed on the line  ensures a unidirectional setup -
modes are propagating from oscillator 1 to 2 and from 2 to the termination.
A possible implementation of an oscillator as the antisymmetric mode
of two lumped circuit resonators is shown in the inset above oscillator 1.
This geometry enables good control of the oscillator-line
coupling needed for the state transfer.
In the symmetric case when $C_1 = C_2$ and $L_1 = L_2$ the antisymmetric
mode $I_1 = I_2$ has a voltage node at point $A$ and therefore does not radiate into the
transmission line.  If the values of the capacitors or inductors are
changed to spoil the symmetry, this mode couples into the line.  }
\label{Implementation}
\end{figure}

 Finally, we   discuss possible realizations of
 state-transfer setups involving time-dependent switching. In the case of
optical cavities, if one is in possession of  switches with switching times much
faster than the round trip travel time of light within the cavity, complete state
transfer could be achieved within one round trip time.  In fact,  a
variety of fast cavity dumping techniques have been developed for pulsed optical systems
in which the switching time is considerably faster than the round trip time
\cite{hook66,zitter67,bridges69,klein70,marcus82,marchetti01}. The state transfer
technique analyzed above is, thus, of most interest in cases where switches fast
compared to the round trip time are not available, such as the case
of optical cavities with short round trip times.  Many cavity dumping
techniques \cite{hook66,marcus82} could be adapted to provide the time-dependent
coupling  Eq. (\ref{15c}) required for oscillator 1 and a switching time that
is fast compared to the ring-down time for oscillator 2.

Since superconducting electronics operating in the microwave frequency range is a
contender for quantum computation, how to achieve fast high fidelity quantum state
transfer between microwave cavities or lumped circuit LC resonators has become an issue
of interest.  Cavity dumping techniques have been employed to produce intense microwave
pulses from low intensity microwave sources  \cite{birx78,alvarez81}.  These
techniques often position the output port at a cavity node.  Coupling to the outside
world is achieved by switching the length of one arm of the cavity so that the output
port is no longer positioned at the node. The inset above oscillator 1 in Fig. \ref{Implementation} shows a
way in which this idea can be implemented in a resonator consisting of lumped circuit
components. If $C_1 = C_2$ and $L_1 = L_2$, the antisymmetric mode of
oscillation, for which $I_1 = I_2,$ has a   voltage node at $A$ and therefore does not radiate
into the transmission line. By changing the value of one or
more of the inductors or capacitors, the symmetry is spoiled, point $A$ will no longer be a node and this
mode will now radiate into the transmission line. The coupling strength is determined
by the size of the shift from symmetry.  A coupling of the form
Eq. (\ref{15c}) can thus be engineered through the use of capacitors and
inductors  having a controllable time-dependent reactance. Time
dependent reactances are generally produced from nonlinear reactors by changing the
operating point of the reactance through the application of bias currents or voltages.
Examples of  a nonlinear reactance that could be used for such an
application are Josephson junctions which function as nonlinear inductors, and varactor
diodes which function as nonlinear capacitors.

To summarize, we have shown that a transfer of an arbitrary quantum state
between two identical harmonic oscillators (optical resonators, electronic
oscillators, etc) coupled to distant ends of a transmission line
can be achieved with a state-independent fidelity that approaches unity  exponentially fast
in the coupling switching-time.

K. J. thanks P. Rabl for helpful discussions.
This work was supported by the Austrian Academy of Sciences,
 Austrian Science Fund and the EuroSQIP network.


\begin{thebibliography}{99}

\bibitem{Cirac}  J. I. Cirac, P. Zoller, H. J. Kimble, and H. Mabuchi,
Phys. Rev. Lett. \textbf{78}, 3221  (1997).

\bibitem{Pellizzari} T. Pellizzari, Phys. Rev. Lett. 79,  5242  (1997).

\bibitem{bose} S. Bose, Phys. Rev. Lett. \textbf{91}, 207901 (2003).

\bibitem{bruder}A. Lyakhov and C. Bruder, New J. Phys. \textbf{7}, 181 (2005).

\bibitem{yung} M.-H. Yung, quant-ph/0603179

\bibitem{de Pasquale} S. Paganelli, F. de Pasquale and G. L. Giorgi, Phys. Rev. A \textbf{74}, 012316 (2006).

\bibitem{Yung networks} M.-H. Yung and S. Bose,  Phys. Rev. \textbf{A} 71, 032310 (2005).


\bibitem{tel}  \emph{Continuous variable teleportation} \cite{Kimble12},
in its single-mode form, aims at a transfer of a harmonic oscillator state.
Its analysis (see e.g. Ref. \cite{MIT}) should thus be applicable to infinite-dimensional Hilbert spaces.
However we note that teleportation differs from the state-transfer considered here
 by the use of a classical information channel.

\bibitem{Kimble12} S. L. Braunstein and H. J. Kimble, Phys. Rev. Lett. \textbf{80}, 869 (1998);
A. Furusawa et al.,  Science \textbf{282}, 706 (1998).


\bibitem{MIT} M. Razavi and J. H. Shapiro, Phys. Rev. A, 73, 042303 (2006).

\bibitem{Denker} B. Yurke and J. S. Denker, Phys. Rev. A \textbf{ 29}, 1419 (1984).

\bibitem{Walls} D. F. Walls and G. J. Milburn, {\it Quantum Optics} (Springer, New York, 1994).

\bibitem{Nielsen} See e.g.,  Eq. (9.60), p. 409, Sec. 9.2.2 in M. A. Nielsen and I. L. Chuang, \emph{Quantum Computation and Quantum Information}
(Cambridge University Press, 2000).

\bibitem{hook66} W. R. Hook, R. H. Dishington, and R. P. Hilberg,
Appl. Phys. Lett. {\bf 9}, 125 (1966).

\bibitem{zitter67} R. N. Zitter, W. H. Steier, and R. Rosenberg,
IEEE J. Quantum Electronics {\bf QE-3}, 614 (1967).

\bibitem{bridges69} T. J. Bridges and P. K. Cheo, Appl. Phys. Lett. {\bf 14},
262 (1969).

\bibitem{klein70} M. B. Klein and D. Maydan, Appl. Phys. Lett. {\bf 16},
509 (1970).

\bibitem{marcus82} S. Marcus, J. Appl. Phys. {\bf 53}, 6029 (1982).

\bibitem{marchetti01}
S. Marchetti, M. Martinelli, R. Simili, M. Giorgi, and
R. Fantoni, Appl. Phys. B {\bf 72}, 927 (2001).


\bibitem{birx78}
D. Birx, G. J. Dick, W. A. Little, J. E. Mercereau,
and D. J. Scalapino, Appl. Phys. Lett. {\bf 32}, 68 (1978).

\bibitem{alvarez81}
R. A. Alvarez et al., IEEE Trans. Mag. {\bf MAG-17}, 935 (1981).
\end{thebibliography}
\end{document}